\documentclass[prl,twocolumn,superscriptaddress,showpacs,amssymb,amsmath,amsfonts,aps]{revtex4}
\usepackage{graphicx}
\usepackage{dcolumn}
\begin{document}
\title{Measurement of the Proton Spin Structure Function $g_1(x,Q^2)$
for $Q^2$ from 0.15 to 1.6 GeV$^2$ with CLAS \\}
%
%

\newcommand*{\AMERICAN }{ American University, Washington, DC 20016}
\affiliation{\AMERICAN}
\newcommand*{\ASU }{ Arizona State University, Tempe, Arizona 85287-1504} 
\affiliation{\ASU } 
\newcommand*{\UCLA }{ University of California at Los Angeles, Los Angeles, 
California  90095-1547} \affiliation{\UCLA } 
\newcommand*{\CMU }{ Carnegie Mellon University, Pittsburgh, Pennsylvania 
15213} \affiliation{\CMU } 
\newcommand*{\CUA }{ Catholic University of America, Washington, D.C. 
20064} \affiliation{\CUA } 
\newcommand*{\SACLAY }{ CEA-Saclay, Service de Physique Nucl\'eaire, F91191 
Gif-sur-Yvette, Cedex, France} \affiliation{\SACLAY } 
\newcommand*{\CNU }{ Christopher Newport University, Newport News, Virginia 
23606} \affiliation{\CNU } 
\newcommand*{\UCONN }{ University of Connecticut, Storrs, Connecticut 
06269} \affiliation{\UCONN } 
\newcommand*{\DUKE }{ Duke University, Durham, North Carolina 27708-0305} 
\affiliation{\DUKE } 
\newcommand*{\EDINBURGH }{ Edinburgh University, Edinburgh EH9 3JZ, United 
Kingdom} \affiliation{\EDINBURGH } 
\newcommand*{\FIU }{ Florida International University, Miami, Florida 
33199} \affiliation{\FIU } 
\newcommand*{\FSU }{ Florida State University, Tallahassee, Florida 32306} 
\affiliation{\FSU } 
\newcommand*{\GWU }{ The George Washington University, Washington, DC 
20052} \affiliation{\GWU } 
\newcommand*{\GLASGOW }{ University of Glasgow, Glasgow G12 8QQ, United 
Kingdom} \affiliation{\GLASGOW } 
\newcommand*{\INFNFR }{ INFN, Laboratori Nazionali di Frascati, Frascati, 
Italy} \affiliation{\INFNFR } 
\newcommand*{\INFNGE }{ INFN, Sezione di Genova, 16146 Genova, Italy} 
\affiliation{\INFNGE } 
\newcommand*{\ORSAY }{ Institut de Physique Nucleaire ORSAY, Orsay, France} 
\affiliation{\ORSAY } 
\newcommand*{\ITEP }{ Institute of Theoretical and Experimental Physics, 
Moscow, 117259, Russia}
\affiliation{\ITEP } 
\newcommand*{\JMU }{ James Madison University, Harrisonburg, Virginia 
22807} \affiliation{\JMU } 
\newcommand*{\KYUNGPOOK }{ Kyungpook National University, Daegu 702-701, 
South Korea} \affiliation{\KYUNGPOOK } 
\newcommand*{\MIT }{ Massachusetts Institute of Technology, Cambridge, 
Massachusetts  02139-4307} \affiliation{\MIT } 
\newcommand*{\UMASS }{ University of Massachusetts, Amherst, Massachusetts  
01003} \affiliation{\UMASS } 
\newcommand*{\MSU }{ Moscow State University, 
Moscow, 119899, Russia} 
\affiliation{\MSU } 
\newcommand*{\UNH }{ University of New Hampshire, Durham, New Hampshire 
03824-3568} \affiliation{\UNH } 
\newcommand*{\NSU }{ Norfolk State University, Norfolk, Virginia 23504} 
\affiliation{\NSU } 
\newcommand*{\OHIOU }{ Ohio University, Athens, Ohio  45701} 
\affiliation{\OHIOU } 
\newcommand*{\ODU }{ Old Dominion University, Norfolk, Virginia 23529} 
\affiliation{\ODU } 
\newcommand*{\PITT }{ University of Pittsburgh, Pittsburgh, Pennsylvania 
15260} \affiliation{\PITT } 
\newcommand*{\RPI }{ Rensselaer Polytechnic Institute, Troy, New York 
12180-3590} \affiliation{\RPI } 
\newcommand*{\RICE }{ Rice University, Houston, Texas 77005-1892} 
\affiliation{\RICE } 
\newcommand*{\URICH }{ University of Richmond, Richmond, Virginia 23173} 
\affiliation{\URICH } 
\newcommand*{\SCAROLINA }{ University of South Carolina, Columbia, South 
Carolina 29208} \affiliation{\SCAROLINA } 
\newcommand*{\UNION }{ Union College, Schenectady, New York 12308}
\affiliation{\UNION }
\newcommand*{\UTEP }{ University of Texas at El Paso, El Paso, Texas 79968} 
\affiliation{\UTEP } 
\newcommand*{\JLAB }{ Thomas Jefferson National Accelerator Facility, 
Newport News, Virginia 23606} \affiliation{\JLAB } 
\newcommand*{\VT }{ Virginia Polytechnic Institute and State University, 
Blacksburg, Virginia   24061-0435} \affiliation{\VT } 
\newcommand*{\VIRGINIA }{ University of Virginia, Charlottesville, Virginia 
22901} \affiliation{\VIRGINIA } 
\newcommand*{\WM }{ College of Willliam and Mary, Williamsburg, Virginia 
23187-8795} \affiliation{\WM } 
\newcommand*{\YEREVAN }{ Yerevan Physics Institute, 375036 Yerevan, 
Armenia} \affiliation{\YEREVAN } 

\author{R.~Fatemi}\affiliation{\VIRGINIA}
\author{A.V.~Skabelin}\affiliation{\MIT}
\author{V.D.~Burkert}\affiliation{\JLAB}
\author{D.~Crabb}\affiliation{\VIRGINIA}
\author{R.~De Vita}\affiliation{\INFNGE}
\author{S.E.~Kuhn}\affiliation{\ODU}
\author{R.~Minehart}\affiliation{\VIRGINIA}
\author{G.~Adams}\affiliation{\RPI}
\author{E.~Anciant}\affiliation{\SACLAY}
\author{M.~Anghinolfi}\affiliation{\INFNGE}
% ...
\collaboration{The CLAS Collaboration}     \noaffiliation

\date{\today}
{\begin{abstract}
 Double-polarization asymmetries for inclusive $ep$ scattering were measured 
 at Jefferson Lab using 2.6 and 4.3 GeV longitudinally polarized electrons
 incident on a longitudinally  polarized NH$_3$ target in the CLAS detector.
 The polarized structure function $g_1(x,Q^2)$ was extracted throughout
 the nucleon resonance region and into the deep inelastic regime, for 
 $Q^2 = 0.15 -1.64~$GeV$^2$. The contributions to the first moment 
 $\Gamma_1(Q^2) = \int g_1(x,Q^2)dx$ were determined up to $Q^2=1.2$ GeV$^2$.
 Using a
 parametrization for $g_1$ in the unmeasured low $x$ regions, the complete
 first moment was estimated over this $Q^2$ region. A rapid change
 in $\Gamma_1$ is observed  for $Q^2 < 1~$GeV$^2$, with a sign change
 near $Q^2 = 0.3~$GeV$^2$, indicating dominant contributions from the
 resonance region. At $Q^2=1.2$ GeV$^2$ our data are below the pQCD
 evolved scaling value.
\end{abstract}
}
\maketitle
Electron scattering has played a long and distinguished role
in the study of nucleon structure.  Inclusive deep inelastic scattering (DIS)
studies \cite{ashman88} revealed the parton constituents (quarks and gluons)
of the nucleon. At low values of the four momentum transfer $Q^2$
the nucleon structure functions depend on both $Q^2$ and on the energy transfer
$\nu=E_0-E'$, where $E_0$ and $E'$ are the initial and final electron energies.
In the asymptotic limit, $Q^2\to\infty$, the scattering is described
by perturbative Quantum Chromodynamics (pQCD) as the absorption of a virtual
photon on a single free quark. In this limit
the structure functions depend only on the Bjorken scaling variable,
$x=Q^2/2M\nu$, where $M$ is the nucleon mass.
  
When both the incident electron and the target nucleon are polarized
(double-polarization) the scattering cross
section depends on two additional functions of $Q^2$ and $\nu$,
$g_1$ and $g_2$. In the
framework of pQCD these also depend only on $x$ as $Q^2\to \infty$. In this
limit $g_1(x)$ has a simple interpretation as the sum over the $x$-dependent 
polarizations of the various quark flavors. A double-polarization experiment
at CERN showed that in the framework of asymptotic QCD only a fraction of
the nucleon spin
could be attributed to the intrinsic spin of the quarks. These measurements
were confirmed by subsequent experiments at CERN, SLAC, and DESY, (see the
reviews \cite{Fil-Ji, Rith-02} and references therein).
Corrections (higher-twist terms) for finite $Q^2$  result in excellent fits
to the extensive data set for $Q^2$ down to the order of 2 GeV$^2$.
It is generally accepted that the intrinsic spin of the quarks accounts
for about 25\% of the nucleon spin, so that other degrees of freedom,
such as gluons and quark orbital angular momentum, must account for the rest.
\medskip

Until recently, only a few double-polarization experiments were carried out
with energies below 25 GeV,  an important example being
 the ground-breaking experiment at SLAC \cite{baum-80} in the late 1970's.
More recently, the spin structure function $g_1$ and its first moment
$\Gamma_1(Q^2)=\int\,g_1(x,Q^2)\,dx$ have become a focus at lower $Q^2$ and in the
resonance region in experiments at SLAC \cite{Abe-98}, 
Hermes \cite{Airapetian-00, Airapetian-03-2}, and
JLab \cite{jlaba-02,Yun-03} to obtain a better understanding
of QCD in the confinement regime and  to study the 
phenomenon of duality between the resonance region and the deep inelastic
region. The
rapidly changing helicity structure of some resonances as a function
of Q$^2$ is expected to have a strong influence on $g_1$.
Double-polarization measurements at lower energies
can therefore indicate where pQCD breaks down and determine
where multi-parton processes and coherent effects due to nucleon resonance
transitions are important.

For inclusive scattering of electrons and protons polarized along the
axis of the electron beam, the double-polarization asymmetry is given by:
\begin{equation}
A_{exp}=\frac{\sigma^{\uparrow\downarrow}-\sigma^{\uparrow\uparrow}}
{\sigma^{\uparrow\downarrow}+\sigma^{\uparrow\uparrow}}
=\sqrt{1-\epsilon^2}~\cos\theta_\gamma 
\left[\frac{A_1 +\eta A_2}{1+\epsilon R}\right],
\label{eqn:pol-asy}
\end{equation}
where $\sigma^{\uparrow\uparrow}$ and $\sigma^{\uparrow\downarrow}$ are
the cross sections for the electron and proton
spins parallel and anti-parallel, respectively.
The factor $\epsilon=[1+2(1+\nu^2/Q^2)\tan^2(\theta/2)]^{-1}$ is the
virtual photon polarization for electron scattering angle $\theta$.
The parameter $R=\sigma_L/\sigma_T$ is the ratio of
the absorption cross sections for longitudinal and transverse
virtual photons. The kinematical factor
 $\eta =\epsilon\sqrt{Q^2}/(E_0-\epsilon~ E')$.
The angle between the virtual photon, $\gamma^*$,
and the beam direction is given by $\theta_\gamma$.

The photon asymmetries $A_1$ and $A_2$ can be written in
terms of the virtual photon absorption cross sections as:
\begin{equation}
 A_1(x,Q^2)=\frac{\sigma^{1/2}_T -\sigma^{3/2}_T}{2\sigma_T},~~~
 A_2(x,Q^2)=\frac{\sigma_{LT}}{\sigma_T},
\label{eqn:a1a2}
\end{equation}
in which $\sigma_T^{1/2}$ and $\sigma_T^{3/2}$ are transverse cross
sections and $\sigma_T$ is half their sum. The superscripts denote the
helicity
of the  $\gamma^* p$  system. The cross section $\sigma_{LT}$ arises from 
longitudinal-transverse interference.
In the nucleon resonance region the relative contributions of helicity
$1/2$ and $3/2$ vary from one resonance to another and depend
strongly on $Q^2$.  
The spin structure function $g_1$ is linearly related to
$A_1$ and $A_2$  by: 
\begin{equation}
 g_1(x,Q^2)=\frac{\nu^2}{Q^2+\nu^2}\left(A_1+\sqrt{\frac{Q^2}{\nu^2}}A_2\right)
F_1(x,Q^2), 
\label{eqn:g1}
\end{equation}
where $F_1$ is a structure function appearing in the
unpolarized electron scattering cross section.

If the contribution from elastic scattering ($x=1$) is excluded,
$\Gamma_1$ vanishes at $Q^2=0$, where, in addition, its slope is constrained
to be negative by
the Gerasimov-Drell-Hearn (GDH) sum rule \cite{GDH1, GDH2} for
absorption of real photons. The validity of the GDH sum rule has been
experimentally tested to better than 10\% by experiments at Mainz \cite{ahrens}
and at ELSA \cite{GDH_collab}. On the other hand, $\Gamma_1$ is known to
be positive at high $Q^2$ \cite{Fil-Ji, Rith-02}.
Therefore, the constraints near $Q^2 =0$ and at high $Q^2$ imply that 
$\Gamma_1$ must change sign at some low $Q^2$, where it is expected to
be dominated by the nucleon resonances \cite{Burkert-93, Drechsel-01}.

At low $Q^2$, meson-baryon dynamics have been
treated in chiral perturbation theory by Ji and Osborne \cite{ji1}
 and by Bernard, Hemmert, and Meissner \cite{bern02} to evolve $\Gamma_1$
to non-zero values of $Q^2$. Badelek, Kwieci\`nski, and Ziaja \cite{Badelek-02}
 have used
the generalized vector dominance model and GDH to make predictions for
$g_1(x)$ at low Q$^2$. At high $Q^2$ the asymptotic value of
$\Gamma_1$ has
been evolved down to $Q^2=1$~GeV$^2$ by using pQCD and the Operator
Product Expansion \cite{DIS-evol}. 
Phenomenological approaches, either with explicit inclusion of resonance
parameters \cite{burkert-ioffe2, Simula-02} or with  general
parametrizations of structure functions \cite{Soffer}, have been used
to cover the entire $Q^2$ range.

The present measurements were carried out with 2.6 and 4.3 GeV longitudinally
polarized electrons incident on a longitudinally polarized target located
at the center of the CLAS detector \cite{Mecking-03}. A toroidal magnetic
field, symmetric about the beam axis, is generated by six super-conducting
coils. The coils separate the detector into six independent spectrometers
that use wire drift chambers for track reconstruction, scintillation counters
for time-of-flight measurements, threshold gas Cerenkov counters, and
lead-scintillator electromagnetic calorimeters. Electrons can be detected
and identified for momenta down to 0.35 GeV/c and for polar angles
from about 8$^\circ$ to 50$^\circ$. The polarization of the beam, which
was measured frequently with a M$\o$ller polarimeter, was typically 70\%.
The beam helicity was flipped at a rate of 1~Hz in a pseudo-random
sequence to minimize systematic effects.

A microwave pumped solid nuclear target \cite{Keith}  using
the method of Dynamic Nuclear Polarization \cite{Abragam} was built
for the CLAS. Its design is similar to those employed
in experiments at SLAC \cite{Abe-98}. Horizontal cylinders, $\sim 1$~cm
long, were packed with ammonia pellets (either $^{15}$NH$_3$ or $^{15}$ND$_3$).
The cylinders were mounted with their axes along the beam line
on a movable ladder, along with an empty cup and a 2.2~mm thick carbon disk.
The target cell was immersed in a liquid He bath at T=1.2~K.
A pair of super-conducting Helmholtz coils coaxial with the beam
generated a uniform 5~T magnetic field in the target volume.
A maximum proton polarization of ~70\% was obtained.
During beam irradiation the proton polarization typically dropped 
gradually to ~40\%,
at which point the target material was removed and replaced in order to 
restore the original polarization.

Charged particles were identified from the measured momentum and
time-of-flight.
The Cerenkov counters and calorimeters were used to reduce the $\pi^-$
contamination in the electron sample to less than 1\%. 
The scattered electrons were binned according to $Q^2$ and the invariant mass 
$W=(M^2+2M\nu-Q^2)^{1/2}$ of the recoiling hadronic system. For each bin the 
polarization asymmetry defined in eqn. \ref{eqn:pol-asy} is:
\begin{equation}
 A_{exp}= C_N C_{ps}\frac{1}{P_eP_t}
\left(\frac{N^{\downarrow\uparrow}-N^{\uparrow\uparrow}}
{N^{\downarrow\uparrow}+N^{\uparrow\uparrow}-N_{bkg}}\right)+A_{RC},
\label{eqn:expasy}
\end{equation}
in which $P_e$ and $P_t$ are the averaged
polarizations of the electrons and protons, respectively; 
$N^{\uparrow\uparrow}$ and $N^{\downarrow\uparrow}$ are
the number of observed electrons normalized to the total incident beam flux 
for parallel and anti-parallel beam and target spins;
and $N_{bkg}$ is the number of electrons
scattered from unpolarized material, which consists
of He, $^{15}$N, target windows, and foils used to isolate vacuum regions.
The factor $C_N=0.98$ corrects
for the contribution of polarized protons in the $^{15}$N \cite{Abe-98}.
The ``pair-symmetric'' correction $C_{ps}$ and the
radiative correction term $A_{RC}$ are discussed below.

The background $N_{bkg}$ accounts for approximately 85\% of the detected
electrons, and it was determined from an analysis of runs using 
the carbon target.  Comparisons of the carbon and ammonia
scattering rates for $W>1.4$ GeV were used to extract the mass
thickness of the ammonia. The known densities and
thicknesses of the materials in the beam-line, along with a parametrization
of the ratio of the free proton and neutron cross sections to account
for the unpaired proton in $^{15}$N, were used to calculate $N_{bkg}$
from the carbon target data.

 Eq.~\ref{eqn:expasy} does not
include corrections for detector acceptance or
efficiency since, except for possible changes in the detector between
ammonia and carbon runs, these cancel in the ratio. Less than 5\%
of the data showed short-term localized changes in the detector. 
The affected regions were removed from the analysis. 

The polarization product $P_eP_t$ for each set of runs
was extracted directly by comparing the measured
asymmetry for elastic scattering to the known elastic $ep$ asymmetry.
Two independent procedures were followed. In the first method,
the background-subtracted asymmetry was obtained in several $Q^2$
bins, checked for consistency, and averaged.
The second method exploited a limited kinematical
range in which both the scattered electron and recoil proton could be
detected. The strict correlation in angle and momentum
for elastic $ep$ scattering distinguishes it 
from inelastic $ep$ scattering and from
quasi-elastic scattering in nitrogen without subtracting a background
spectrum. The two methods were consistent with each other. A typical
value was $P_eP_t=(36\pm 1)$\%.
\begin{figure}[h]
\includegraphics[scale=0.48]{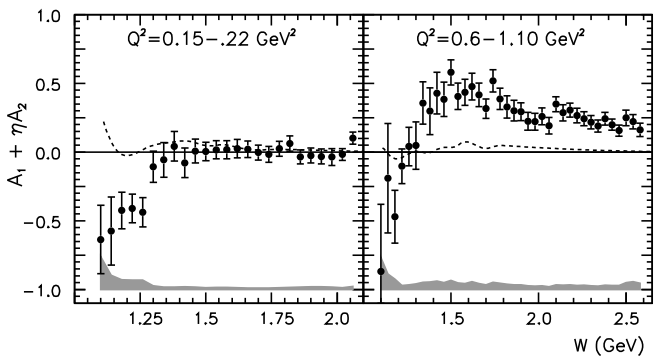}
\caption{The photon asymmetry $A_1+\eta A_2$ vs. the invariant hadronic
mass $W$(GeV) for the proton in two $Q^2$ bins. The lower $Q^2$ data were
obtained with a beam energy of 2.6 GeV and the higher with 4.3 GeV. The
dashed curves show our model estimates of $\eta A_2$. The error
bars denote statistical uncertainties and the shaded bands near the bottom of
the plot
indicate the magnitude of the systematic uncertainties (1 $\sigma$).
\label{fig:a1na2}}
\end{figure}

A ``pair-symmetric'' correction $C_{ps}$  was
made for the contribution of $\pi^0\to e^+e^-\gamma$ to the inclusive  
electron sample. It was determined by measuring the $e^+$ rate with
opposite torus current. Since the correction rises very sharply at
high $W$ (low scattered electron energy), we analyzed data only in the
kinematic region where the pair-symmetric background was less than
10\% of the total electron rate. 

Radiative corrections were applied to the experimental asymmetries
using the code RCSLACPOL \cite{Abe-98}, developed at SLAC and based
on the approach of Kuchto and Shumeiko \cite{kukh-83} for the internal
corrections, and by Tsai \cite{tsai-74} for external 
corrections. We used parametrizations of the world data on unpolarized
and polarized structure functions (including our own preliminary
asymmetry data) and elastic form factors as input
for the radiative correction code. Details of this model will
be given in a longer paper.

\begin{figure}[h]
\includegraphics[scale=0.46]{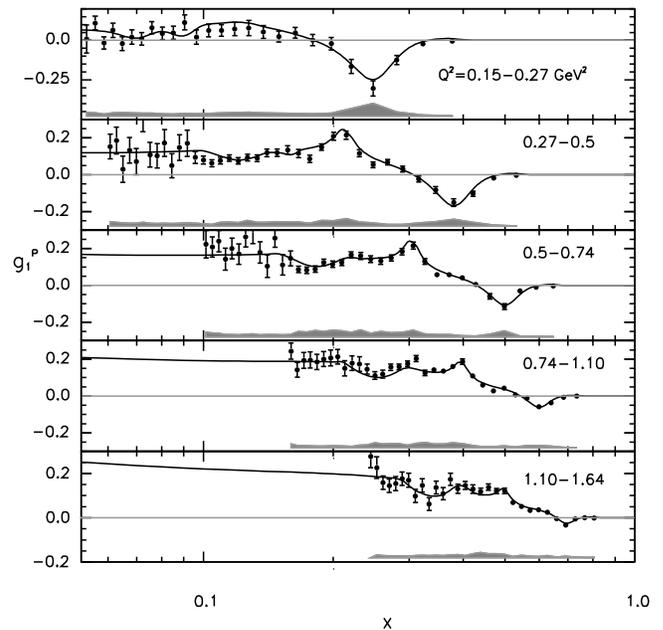}
\caption{The polarized structure function $g_1(x)$ vs. $x$ for the proton
for five $Q^2$ bins.
The solid lines show the parametrization described in the text. The error
bars denote statistical uncertainties and the shaded area
at the bottom of each plot indicates the magnitude of the systematic 
uncertainties (1 $\sigma$).
\label{fig:g1x}}
\end{figure}

The model used for radiative corrections was also used to calculate
 values for $A_1$,
$A_2$, $g_1$, $g_2$, $F_1$, $F_2$, and $R$ over the measured region of
$Q^2$ and $x$. Using the calculated value of $R$, Eq.~\ref{eqn:pol-asy}
was used to obtain $A_1+\eta A_2$ from the corrected experimental asymmetry
defined by Eq.~\ref{eqn:expasy}. The results for two representative $Q^2$
bins are shown in Fig.~\ref{fig:a1na2}.
The model values of $\eta A_2$ were generally small, as can
be seen from the dashed curves in
Fig.~\ref{fig:a1na2}. These calculated values were used
to extract $A_1$ from the measured $A_{exp}$. Finally, we used
the model for $F_1$ and $A_2$, along with Eq.~\ref{eqn:g1}, to extract $g_1$,
which is dominated by $A_1$. The results for $g_1$ for the proton for
five $Q^2$ bins are shown in Fig.~\ref{fig:g1x}.
The solid curve shows $g_1$ calculated from the data parametrization
used in RCSLACPOL. 

\begin{figure}[h]
\includegraphics[scale=0.45]{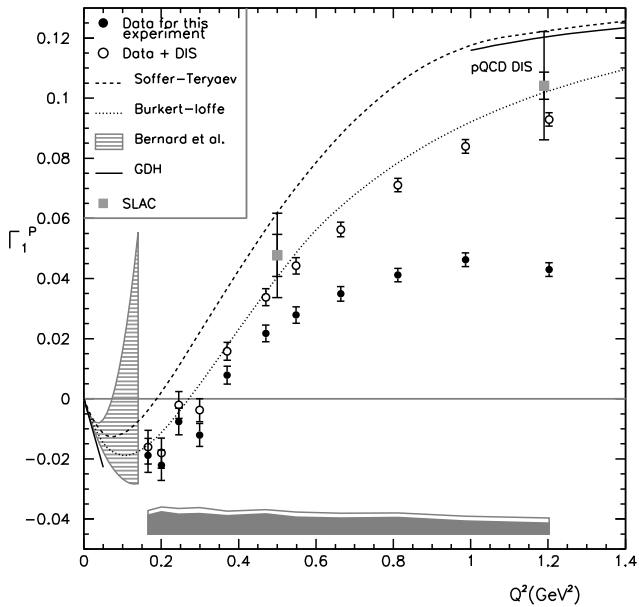}
\caption{The first moment $\Gamma_1$ 
 vs. $Q^2$ for the proton. The closed circles are obtained
by integrating over our measurements of $g_1$. The open circles
are obtained by using the inclusive scattering model to extend the integral
down to $x=10^{-5}$. The error bars indicate statistical uncertainties. The dark
shaded region at the bottom shows the estimated systematic
uncertainties~(1 $\sigma$)
for the measured points. The line above the shaded region indicates
the systematic uncertainty for the open circles.
Measurements from SLAC \cite{Abe-98} are shown as shaded squares with the
double error bar indicating statistical uncertainty and (statistical + 
systematic) uncertainties. See the text for an explanation of the curves.}
\label{fig:g1int3}
\end{figure}

We can integrate $g_1(x,Q^2)$ from the lowest measured value of $x$
(with a cut-off ranging from $W$=2 to 2.6~GeV) up
to $x=1$ at each $Q^2$ to obtain the contribution of our data to the
integral $\Gamma_1$ (the contribution of elastic scattering is excluded).
The results are shown as the closed circles in Fig.~\ref{fig:g1int3}.
The full value of the integral was estimated by using the
parametrization for $g_1$ to estimate the contribution from $x=10^{-5}$
up to the threshold of our measurements. These extended integrations are
shown as open circles. The plot is cut off
at $Q^2=1.2$ GeV$^2$ since at higher values more than half the integral
is due to the unmeasured region of $x$. 
 The slope at $Q^2=0$
required by the GDH sum rule is indicated by a straight line.
Calculations using the light baryon ChPT formulation of
Bernard et al. \cite{bern02} are indicated by a shaded band at 
$Q^2<0.2$~GeV$^2$
with boundaries that reflect uncertainties in resonance parameters.
The pQCD evolution of the deep inelastic
scattering measurements to ${\cal{O}}(\alpha_s^3$) \cite{DIS-evol}
is shown as a line at high $Q^2$. The calculations
of Soffer and Teryaev \cite{Soffer} and
Burkert and Ioffe \cite{burkert-ioffe2}
in the intermediate region are also shown.  A linear fit to 
the five points from $Q^2$=0.20 to 0.47 GeV$^2$ yields a zero crossing for
$\Gamma_1$  at $Q^2=0.29\pm 0.03$~GeV$^2$, where the error includes
only the statistical uncertainty and the estimated uncertainty in the DIS
contribution added in quadrature. The zero crossing indicates the
transition to a distance scale where non-partonic contributions such
as resonance excitations are  dominant. Our results for $\Gamma_1$($Q^2$)
lie well below the predictions from the pQCD evolution from DIS. They
are in better agreement with the model calculations of
Ref.~{\cite{burkert-ioffe2}} that include s-channel baryon resonance
excitations explicitly. Thus, we think that it is likely that the
lack of explicit inclusion of the resonance contributions in the pQCD
evolution gives rise to the discrepancy.

The estimated systematic uncertainties (1 $\sigma$) are indicated by
the shaded band at the bottom of each plot in the figures.
The systematic uncertainty is dominated by the parametrizations
of  $A_2$, $F_1$ and $R$, which constitute 75\%  of the total
uncertainty at low $Q^2$ and 50\% at high $Q^2$. The 
uncertainty is estimated by using alternative parametrizations, 
as well as by setting $R$ and A$_2$ to extreme values. The radiative
corrections, which also incorporate these models, constitute
20\% of the total systematic uncertainty at low $Q^2$ and 5\% at high $Q^2$.
The remaining uncertainty arises largely from livetime calculations and
the removal of the nuclear contributions from
the ammonia spectra. The systematic uncertainty of the DIS extrapolation 
for Fig.~\ref{fig:g1int3} was estimated by using three different
parametrizations for $g_1$ in the low $x$ region,
Simula \cite{Simula-02}, and the previously mentioned model fitted
to world data before 1999 and 2000, respectively.

In summary, our measured asymmetries for inclusive
scattering of 2.6 and 4.3 GeV polarized electrons on polarized protons
have been used to extract the structure function $g_1(x,Q^2)$ for $Q^2$
from 0.15 to 1.64 GeV$^2$. The first moment $\Gamma_1$ depends strongly on
$Q^2$,  with its sign changing near $Q^2= 0.3$ GeV$^2$. 

We would like to acknowledge the outstanding efforts of the staff of the
Accelerator, the Target Group, and the Physics Divisions at JLab who
made this experiment possible. This work was supported in part by
the U.S. Department of Energy, the U.S. National Science Foundation,
the Istituto Nazionale di Fisica Nucleare, the  French Centre National de
la Recherche Scientifique, the French Commissariat \`{a} l'Energie Atomique,
and the Korean Science and Engineering Foundation. The Southeastern
Universities Research Association (SURA) operates the Thomas Jefferson National
Accelerator Facility for the United States Department of Energy under
contract DE-AC05-84ER40150.

\end{document}